Nanoscale Research Letters
a SpringerOpen Journal

NANO EXPRESS                                                                                    Open Access

# AFM-assisted fabrication of thiol SAM pattern with alternating quantified surface potential

Bradley Moores[1], Janet Simons[2], Song Xu[3], Zoya Leonenko[1,2*]

## Abstract

Thiol self-assembled monolayers (SAMs) are widely used in many nano- and bio-technology applications. We report a new approach to create and characterize a thiol SAMs micropattern with alternating charges on a flat gold-coated substrate using atomic force microscopy (AFM) and Kelvin probe force microscopy (KPFM). We produced SAMs-patterns made of alternating positively charged, negatively charged, and hydrophobic-terminated thiols by an automated AFM-assisted manipulation, or nanografting. We show that these thiol patterns possess only small topographical differences as revealed by AFM, and distinguished differences in surface potential (20-50 mV), revealed by KPFM. The pattern can be helpful in the development of biosensor technologies, specifically for selective binding of biomolecules based on charge and hydrophobicity, and serve as a model for creating surfaces with quantified alternating surface potential distribution.

## Background

Thiol self-assembled monolayers (SAMs) are promising for many nano- and bio-technology applications as they offer a reliable method to produce surfaces with desirable properties. These properties can be used for specific and non-specific binding of biomolecules and nanoparticles and, therefore, can serve as useful templates for nano- and micro-fabrication. The first systematic study of thiol chemicals was reported by Zisman and co-authors [1], and has since been investigated by many researchers, including a detailed review by Chechik et al [2]. SAMs can be defined as "molecular assemblies that are formed spontaneously be the immersion of an appropriate substrate into a solution of an active surfactant in an organic solvent" [3]. Thiols are a perfect type of such surfactant as they consist of a surface-active sulfur group that binds to the metal surface, a hydrocarbon chain of various lengths that defines the packing of the monolayer, and a functional group at the end that determines the functional properties of the formed SAM film. When metallic surfaces such as gold, platinum, or silver are exposed to thiols dissolved in organic solvent, a bond is formed between the thiol's active sulfur group and metal atoms of the surface, which is characterized by a shared pair of electrons. Uniform monolayer coverage can be created on flat metallic surfaces using procedures empirically determined for each thiol type, involving factors such as incubation time, solvent, and concentration [4,5].

With the invention of scanning probe microscopy and other nanoscale characterization techniques, much interest has been created in nanoscale fabrication for nanoelectronics and biosensing. The advantage of sensing on the nanoscale using miniaturized devices created a demand in producing thiol SAMs of a repeated pattern, which can be used as biosensing platforms. Many nanopatterning techniques require electron-beam or photo-lithography in vacuum environments [6], using a polymer mask [7,8], or stamping approaches [9], and cannot produce patterns on the nanoscale. The atomic force microscopy (AFM)-based nanopatterning technique simply involves using an atomic force microscope, where the AFM probe is used as a sharp stylus to scratch the thiols from the surface. The force applied by the AFM probe can easily disturb the sulfur bond between the thiols and metal surface. This approach has been demonstrated by producing simple defects in thiol monolayers [10]. This opened the development of a new nanografting method for patterning SAMs with nanometer precision [10]. In this study, we used the new nanografting method to produce a pattern by mechanically substituting one thiol with another using an AFM probe in the solution of the second thiol. The scratched squares

* Correspondence: zleonenk@uwaterloo.ca
[1]Department of Physics and Astronomy, University of Waterloo, 200 University Avenue West, Waterloo, ON N2L 3G1, Canada.
Full list of author information is available at the end of the article produced by the AFM probe were immediately filled by the second thiol present in solution due to the higher chemical concentration. This procedure makes it possible to create an alternating charge pattern composed of two different thiols. We used the AFM to characterize surface morphology and Kelvin probe force microscopy (KPFM) to characterize the surface potential of the produced pattern.

## Results and discussion
### Nanografting

Uniform surface coverage with one thiol was created by incubating a gold surface for 24-72 h in this thiol solution. After incubation, the sample was exposed to a second thiol solution in an AFM liquid cell. The AFM probe, with a spring constant of 5 N/m, was inserted into the liquid cell and was used to scratch squares of defined dimension, varying from 10 by 10 nm to 10 by 10 μm in contact mode. The force applied was just high enough to remove thiol molecules from the surface (approximately between 10 and 50 nN), thus leaving the gold undamaged. We have performed scratching at high speeds (10 lines/s) in order to reduce thermal drift and decrease the time required to create a pattern. The number of lines per square depends on the tip geometry, but we found for 10 μm$^2$ squares at least 512 lines were required. The scratched squares produced by the AFM probe were immediately filled by thiol 2 present in solution due to the higher concentration of this thiol. The second thiol must have a higher affinity for the metallic surface to replace the first thiol removed from the surface by AFM probe. This process makes it possible to create an alternating charge pattern, composed of two different thiols. AFM and KPFM were used to characterize the pattern in terms of topography and surface potential.

We first incubated a solution of $CH_3$-terminated thiol molecules on gold-coated glass for 24 h. Figure 1a shows an AFM topography image of this thiol SAM in air. We applied a three-step nanografting method [11] to produce a pattern. First, AFM was used to image a previously formed monolayer (matrix SAM) in a solution with another thiol (COOH-terminated thiol). Second, the tip was positioned into a selected spot to start a programmed scratching of defined areas. The scratching was performed with a higher load than the threshold for thiol 1 ($CH_3$-terminated) displacement [12]. During the scratching, the AFM probe removed matrix thiol 1 and produced bare gold squares exposed to thiol 2 (COOH-terminated) solution (nanoshaving) [13].

### Surface potential of thiol SAM pattern

Figure 2a shows AFM topography image of the two-thiols pattern, created by substituting thiol 1 ($CH_3$-terminated thiol) with thiol 2 (COOH-terminated thiol).

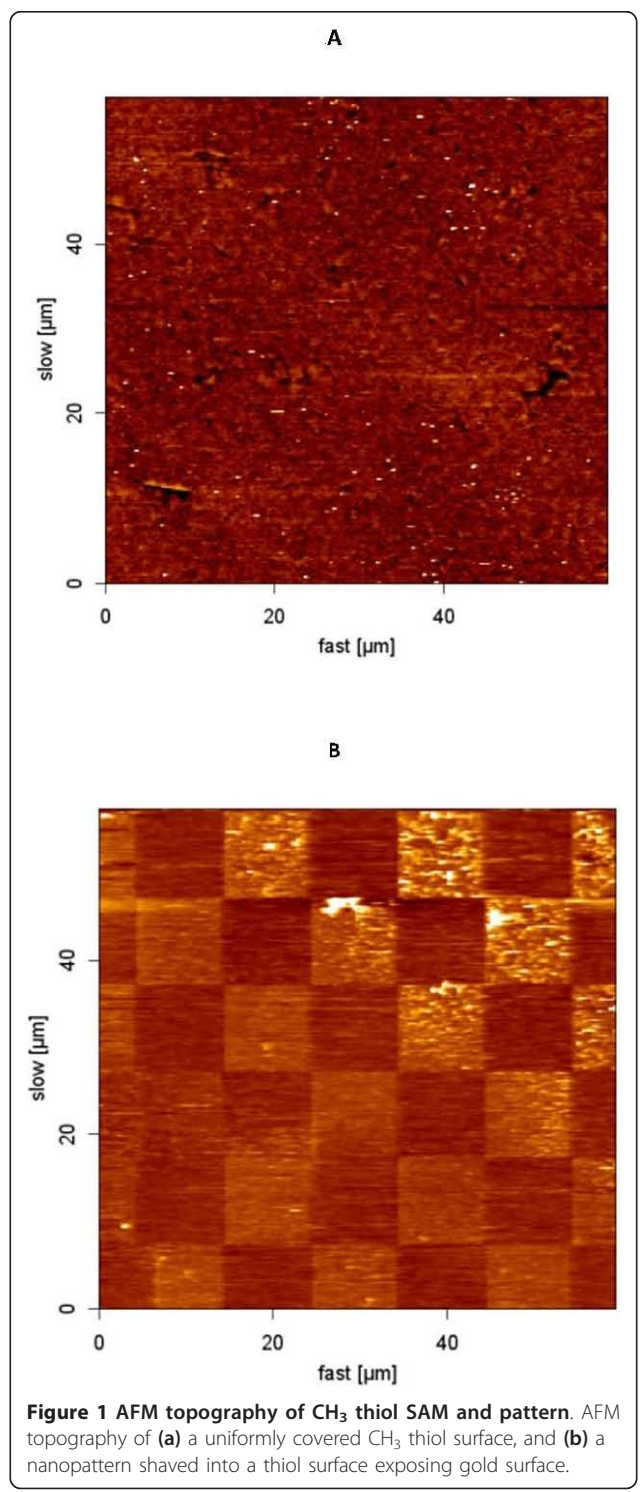

Figure 1 AFM topography of $CH_3$ thiol SAM and pattern. AFM topography of (a) a uniformly covered $CH_3$ thiol surface, and (b) a nanopattern shaved into a thiol surface exposing gold surface.

Topography does not show much contrast as the two thiols do not differ significantly in height. The cross-section plot for topography image shows flat profile, with exception of few impurities. Figure 2b shows a surface potential map, obtained with KPFM and reveals a pronounced difference (20 mV) in surface potential on the border of two thiols (cross-section plot, Figure 2d).

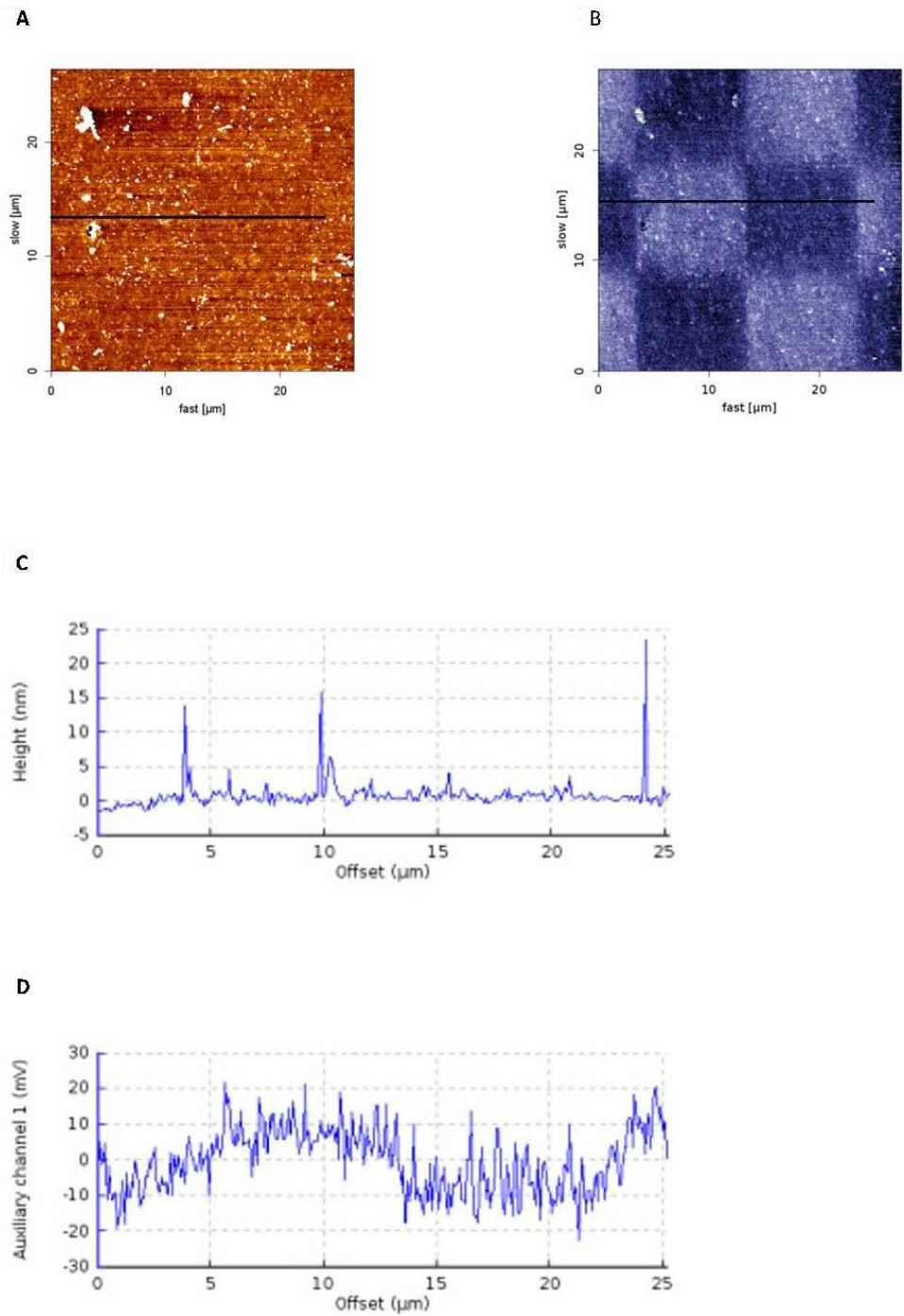

Figure 2 **AFM and KPFM of CH$_3$/COOH thiol pattern**. Nanopattern **(a)** topography and **(b)** KPFM produced using CH$_3$ and COOH thiols. **(c,d)** show cross-sections of the topography and KPFM, respectively.

## Conclusions

In summary, we showed that thiol SAM pattern with chemical functionality and desired surface potential differences can be created using AFM-based nanografting method. In addition, we demonstrated for the first time that small differences in surface potential maps associated with organic thiol patterns can be resolved by KPFM in mV range. Such patterns with controlled differences in surface potential can be useful in nano- and bio-technology applications and to study interactions of charged species, such as nanoparticles and macromolecular ions with non-uniformly charges surfaces.

## Methods
### Chemicals and sample preparation
Decanethiol, cysteamine hydrochloride, 3-mercaptopropionic acid, and HPLC grade ethanol were purchased from Sigma-Aldrich Chemical Co. (St. Louis, MO, USA). These chemicals produce $CH_3$, $NH_2$, and COOH-terminated surfaces, respectively. All chemicals were used as received with no further purification. Thiols were dissolved in ethanol at 5 mM concentration.

### Thiol SAM preparation
Gold-coated mica slides were purchased from Agilent Technologies, Inc. (Santa Clara, CA, USA). Before use, these gold surfaces were glued to clean glass cover slips using Epo-Tek 377 glue from Epo-Tek, Inc. (Billerica, MA, USA), which was cured at 150°C for 1 h [14]. The mica slide was removed from the "sandwich" substrate, leaving the glass with attached gold thin film, revealing the atomically flat gold side. The exposed gold was imaged to confirm atomically flat topography. The gold surfaces were then incubated in an appropriate 5 mM thiol solution in ethanol for 24-72 h to obtain uniform SAM surface coverage.

### Atomic force microscopy
AFM uses a sharp probe over a sample surface and allows for imaging the topographical features at the nanoscale. Two common modes of operation are contact mode and intermittent contact mode. Thiol-modified surfaces were imaged in intermittent contact mode with a JPK Nanowizard II atomic force microscope. In intermittent contact mode, the tip is oscillated at the resonant frequency of the cantilever, and a feedback loop maintains constant amplitude over the entire image to insure the gentle imaging conditions. The probes used were Nanoworld NCH tips with a resonant frequency of approximately 338 kHz and 42 N/m spring constant. In contact mode of imaging, the tip usually lightly touches the surface and is moved up and down with the topographical features of the sample. With increased force the probe can interact strongly with the surfaces and remove soft matter from the surface. This approach was used for nanografting. Automated patterning was achieved by programming the JPK AFM imaging software to scratch a square of defined size and then move to the next defined location. Alternating this process produces the pattern with the size of few $nm^2$ to few $\mu m^2$.

### Kelvin probe force microscopy
KPFM is an extension of AFM that provides the ability to map the surface potential in addition to imaging sample topography [15-17]. KPFM measures the surface potential by eliminating the electrostatic interactions between the tip and sample by applying a DC bias. This DC bias is tuned by a feedback loop that monitors mechanical oscillations induced in the tip due to an AC voltage (1 V) applied to the tip or sample. KPFM images were recorded using lift mode (also known as hover mode) operation. In lift mode KPFM, the topography of the sample is measured during the trace scan without an applied potential. During the retrace of the same line, the tip follows the topography measured during the trace pass but offset 50 nm above the surface, and an AC and DC voltage is applied between the tip and sample to nullify the electrostatic interactions. Increasing the tip-sample separation by 50 nm eliminates the possibility of cross talk between the topography and surface potential measurements.

Surface potential images were recorded in air using Nanoworld NCH cantilevers with a JPK Nanowizard II AFM in a hover mode KPFM. The gold substrates were grounded to eliminate sample charging.

### Abbreviations
AFM: atomic force microscopy; KPFM: Kelvin probe force microscopy; thiols: SAMs (self-assembled monolayers).

### Acknowledgements
The authors acknowledge technical support from JPK Instruments, Germany, and Agilent Technologies, USA. The authors acknowledge financial support from Natural Science and Engineering Council of Canada (NSERC), Canadian Foundation of Innovation (CFI), Ontario Research Fund (ORF), as well as Waterloo Institute for Nanotechnology (WIN) Graduate Scholarship Award to B. Moores.

### Author details
[1]Department of Physics and Astronomy, University of Waterloo, 200 University Avenue West, Waterloo, ON N2L 3G1, Canada. [2]Department of Biology, University of Waterloo, 200 University Avenue West, Waterloo, ON N2L 3G1, Canada. [3]Agilent Technologies, 4330 W. Chandler Blvd. Chandler, AZ 85226, USA.

### Authors' contributions
BM and JS carried out the thiol SAM preparation and nanografting experiments, participated in the manuscript draft preparation. BM carried out KPFM imaging. SX participated in nanografting experiments and participated in the manuscript draft preparation. ZL conceived of the study, participated in its design and coordination and finished the final draft of the manuscript. All authors read and approved the final manuscript.

### Competing interests
The authors declare that they have no competing interests.